\documentclass[conference]{IEEEtran}

\usepackage{bm}% bold math
\usepackage{tikz}
\usetikzlibrary{arrows}
\ifCLASSINFOpdf
\else
\fi
\begin{document}
\title{A Pulse-Gated, Predictive Neural Circuit}

\author{\IEEEauthorblockN{Yuxiu Shao\IEEEauthorrefmark{1}, Andrew T. Sornborger\IEEEauthorrefmark{2}, Louis Tao\IEEEauthorrefmark{3}}\\
\IEEEauthorblockA{\IEEEauthorrefmark{1}\small Center for Bioinformatics, National Laboratory of Protein Engineering and Plant Genetic Engineering,\\ College of Life Sciences, Peking University, Beijing, China\\
Email: shaoyx@mail.cbi.pku.edu.cn}\\
\IEEEauthorblockA{\IEEEauthorrefmark{2} \small Department of Mathematics, University of California, Davis, USA\\
Email: ats@math.ucdavis.edu}\\
\IEEEauthorblockA{\IEEEauthorrefmark{3}\small Center for Bioinformatics, National Laboratory of Protein Engineering and Plant Genetic Engineering,\\ College of Life Sciences, and Center for Quantitative Biology, Peking University, Beijing, China\\
Email: taolt@mail.cbi.pku.edu.cn\\
ATS and LT are corresponding authors.}}
\maketitle

\begin{abstract}
Recent evidence suggests that neural information is encoded in packets and may be flexibly routed from region to region.
We have hypothesized that neural circuits are split into sub-circuits where one sub-circuit controls information propagation via pulse gating and a second sub-circuit processes graded information under the control of the first sub-circuit. Using an explicit pulse-gating mechanism, we have been able to show how information may be processed by such pulse-controlled circuits and also how, by allowing the information processing circuit to interact with the gating circuit, decisions can be made. Here, we demonstrate how Hebbian plasticity may be used to supplement our pulse-gated information processing framework by implementing a machine learning algorithm. The resulting neural circuit has a number of structures that are similar to biological neural systems, including a layered structure and information propagation driven by oscillatory gating with a complex frequency spectrum.
\end{abstract}


% \fi
\IEEEpeerreviewmaketitle


\section{Introduction}
A considerable experimental literature indicates that 1) oscillations of various frequencies are important for the modulation of interactions in neural systems \cite{AzouzGray2000,FriesEtAl2008,MarkowskaEtAl1995}, 2) each individual pulse that makes up an oscillation may be implicated in the parallel transfer of a packet of information \cite{Fries2005,SalinasSejnowski2000,pmid17185420,pmid26507295}, and 3) neural systems exist for the control of information propagation \cite{pmid21267396}. We have begun to build a neural information processing framework based on the pulse-gated propagation of graded (amplitude dependent) information \cite{SornborgerWangTao,pmid27310184}. A key question that arises in understanding information processing in the brain is: How may neural plasticity be used to form computational circuits?

In this paper, we explore this question by creating a predictive neural circuit based on the pulse-gated control of firing rate information. We outline a set of sub-circuits responsible for sub-computations needed to estimate the process coefficients of an autoregressive ($AR$) learning circuit. We then combine the sub-circuits to demonstrate that a neural system can use Hebbian learning in concert with pulse-gating to predict an $AR$ process.

\section{Methods}

\subsection{Autoregressive Processes}
AR and moving-average (MA) processes are the two principle linear models that are used to make statistical predictions \cite{PanditWu,PercivalWalden}. AR models are filters of length $n$ on a time series and we will consider these here. The object of this work is to implement an algorithm for predicting an AR process in a substrate of neurons. Below, we denote the filter by $\mathbf{a}$, the individual process covariances by $\sigma_i$, and the process variance-covariance matrix by $\Sigma$.

An AR($n$) process for a zero-mean random variable, $x_t$, is defined $$x_t = \sum_{i=1}^n a_i x_{t-i} + \epsilon_t \; .$$

To find the filter, we need to solve the Yule-Walker equations, which arise from the covariance structure of the process as follows:
\begin{eqnarray}
\langle x_t x_t \rangle & = & \sum_{i=1}^n a_i \langle x_{t-i} x_t \rangle + \langle \epsilon_t x_t \rangle \nonumber \\
\langle x_t x_{t-1} \rangle & = & \sum_{i=1}^n a_i \langle x_{t-i} x_{t-1} \rangle \nonumber \\
& \vdots & \nonumber \\
\langle x_t x_{t-n} \rangle & = & \sum_{i=1}^n a_i \langle x_{t-i} x_{t-n} \rangle \nonumber 
\end{eqnarray}
where $\langle \rangle$ denotes an expectation value over $t$. Defining $\sigma_i = \langle x_{t-i} x_t \rangle = \langle x_t x_{t-i} \rangle$, from the second through $n$'th equations above, we have
\begin{equation}
   \boldsymbol{\sigma} = \Sigma \mathbf{a} \; , \nonumber
\end{equation}
where $\boldsymbol{\sigma} = (\sigma_1, \dots, \sigma_n)$,
\begin{equation}
   \Sigma = \left[ \begin{array}{cccc} \sigma_0 & \sigma_1 & \dots & \sigma_{n-1} \\
                                                      \sigma_1 & \sigma_0 & \ddots & \\
                                                      \vdots & \ddots & \ddots & \\
                                                      \sigma_{n-1} & & & \sigma_0 \end{array} \right] \; , \nonumber
\end{equation}
and $\mathbf{a} = (a_1, \dots, a_n)$.

\subsection{Push-Me Pull-You Neuron Pairs}

In our AR process, $x_t$, since we are considering an implementation in a neural circuit where firing rates encode information, the input to our system will be positive semi-definite. Both negative and positive values of $x_t - m_0$, where $m_0$ is the mean, must be able to be represented by the circuit. We therefore define neuron pairs where one neuron of the pair represents positive values (i.e. the amplitude above the mean) and the other neuron represents negative values (i.e. the absolute value of the amplitude below the mean). At any given moment, only one neuron will encode non-zero amplitude in such a pair. In order to read in and encode positive and negative values of the input in a pair of neurons, we let $m_0 = \langle x_t \rangle$, $$\tau \frac{dm_1}{dt} = -m_1 + \left[ x - m_0 + p(t) - g_0 \right]^+$$ and $$\tau \frac{dm_2}{dt} = -m_2 + \left[ m_0 - x + p(t) - g_0\right]^+ \; ,$$
where $p(t)$ is a square pulse of duration $T$. Here, $g_0$ is a threshold value and we assume that $|x - m_0|$ is less than the amplitude of the pulse, $p(t)$, so neuron $1$ (firing rate $m_1$) will only fire when the input is simultaneously pulsed and above the mean and, similarly, neuron $2$ will only fire when the input is simultaneously pulsed and below the mean. We call such a pair of neurons a push-me pull-you (PMPY) pair. Additionally, PMPY amplitudes may be exactly propagated along a chain via pulse-gating, and we will also designate pairs along the chain as PMPY pairs.

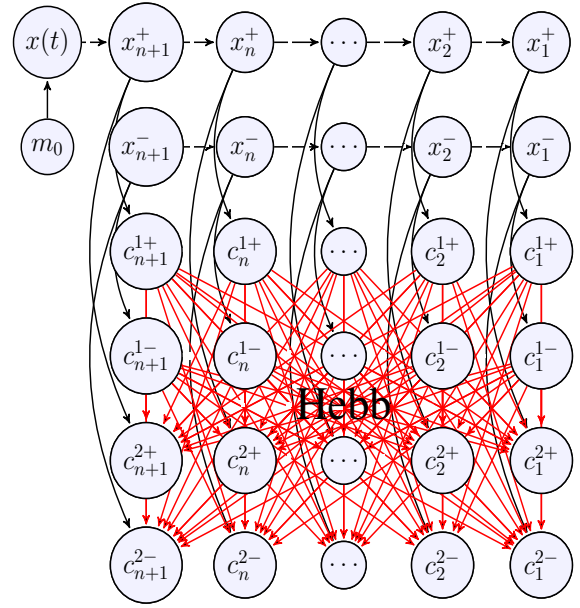
\begin{figure}
\centering
\resizebox{3in}{3.15in}{

\begin{tikzpicture}[->,>=stealth',shorten >=1pt,auto,node distance=1.9cm,
  thick,main node/.style={circle,fill=blue!5,draw,
  font=\sffamily\Large\bfseries,minimum size=6mm},scale=0.25]

  \node[main node] (xp) {$x_{n+1}^+$};
  \node[main node] (xn) [below of=xp] {$x_{n+1}^-$};
  \node[main node] (xt) [left of=xp] {$x(t)$};
  \node[main node] (m)  [left of=xn] {$m_0$};
  \node[main node] (c1p) [below of=xn] {$c_{n+1}^{1+}$};
  \node[main node] (c1n) [below of=c1p] {$c_{n+1}^{1-}$};
  \node[main node] (c2p) [below of=c1n] {$c_{n+1}^{2+}$};
  \node[main node] (c2n) [below of=c2p] {$c_{n+1}^{2-}$};
  \node[main node] (xnm1p) [right of=xp] {$x_{n}^+$};
  \node[main node] (xnm1n) [right of=xn] {$x_{n}^-$};
  \node[main node] (c1nm1p) [right of=c1p] {$c_{n}^{1+}$};
  \node[main node] (c1nm1n) [right of=c1n] {$c_{n}^{1-}$};
  \node[main node] (c2nm1p) [right of=c2p] {$c_{n}^{2+}$};
  \node[main node] (c2nm1n) [right of=c2n] {$c_{n}^{2-}$};
  \node[main node] (xdotsp) [right of=xnm1p] {$\dots$};
  \node[main node] (xdotsn) [right of=xnm1n] {$\dots$};
  \node[main node] (c1dotsp) [right of=c1nm1p] {$\dots$};
  \node[main node] (c1dotsn) [right of=c1nm1n] {$\dots$};
  \node[main node] (c2dotsp) [right of=c2nm1p] {$\dots$};
  \node[main node] (c2dotsn) [right of=c2nm1n] {$\dots$};
  \node[main node] (x2p) [right of=xdotsp] {$x_{2}^+$};
  \node[main node] (x2n) [right of=xdotsn] {$x_{2}^-$};
  \node[main node] (c12p) [right of=c1dotsp] {$c_{2}^{1+}$};
  \node[main node] (c12n) [right of=c1dotsn] {$c_{2}^{1-}$};
  \node[main node] (c22p) [right of=c2dotsp] {$c_{2}^{2+}$};
  \node[main node] (c22n) [right of=c2dotsn] {$c_{2}^{2-}$};
  \node[main node] (x1p) [right of=x2p] {$x_{1}^+$};
  \node[main node] (x1n) [right of=x2n] {$x_{1}^-$};
  \node[main node] (c11p) [right of=c12p] {$c_{1}^{1+}$};
  \node[main node] (c11n) [right of=c12n] {$c_{1}^{1-}$};
  \node[main node] (c21p) [right of=c22p] {$c_{1}^{2+}$};
  \node[main node] (c21n) [right of=c22n] {$c_{1}^{2-}$};

  \path[every node/.style={font=\sffamily\small,
  		fill=white,inner sep=1pt}]
  	% Right-hand-side arrows rendered from top to bottom to
  	% achieve proper rendering of labels over arrows.
    (xp) edge [bend right=25] node[left=1mm] {} (c1p)
         edge [bend right=25] node[left=1mm] {} (c2p)
         edge [bend right=0] node[left=1mm] {} (xnm1p)
    (xn) edge [bend right=25] node[left=1mm] {} (c1n)
         edge [bend right=25] node[left=1mm] {} (c2n)
         edge [bend right=0] node[left=1mm] {} (xnm1n);
  \path[every node/.style={font=\sffamily\small,
  		fill=white,inner sep=1pt}]
  	% Right-hand-side arrows rendered from top to bottom to
  	% achieve proper rendering of labels over arrows.
    (xt) edge [bend right=0] node[left=1mm] {} (xp)
    (m) edge [bend right=0] node[left=1mm] {} (xt);
  \path[every node/.style={font=\sffamily\small,
  		fill=white,inner sep=1pt}]
  	% Right-hand-side arrows rendered from top to bottom to
  	% achieve proper rendering of labels over arrows.
    (xnm1p) edge [bend right=25] node[left=1mm] {} (c1nm1p)
         edge [bend right=25] node[left=1mm] {} (c2nm1p)
         edge [bend right=0] node[left=1mm] {} (xdotsp)
    (xnm1n) edge [bend right=25] node[left=1mm] {} (c1nm1n)
         edge [bend right=25] node[left=1mm] {} (c2nm1n)
         edge [bend right=0] node[left=1mm] {} (xdotsn);
  \path[every node/.style={font=\sffamily\small,
  		fill=white,inner sep=1pt}]
  	% Right-hand-side arrows rendered from top to bottom to
  	% achieve proper rendering of labels over arrows.
    (xdotsp) edge [bend right=25] node[left=1mm] {} (c1dotsp)
         edge [bend right=25] node[left=1mm] {} (c2dotsp)
         edge [bend right=0] node[left=1mm] {} (x2p)
    (xdotsn) edge [bend right=25] node[left=1mm] {} (c1dotsn)
         edge [bend right=25] node[left=1mm] {} (c2dotsn)
         edge [bend right=0] node[left=1mm] {} (x2n);
  \path[every node/.style={font=\sffamily\small,
  		fill=white,inner sep=1pt}]
  	% Right-hand-side arrows rendered from top to bottom to
  	% achieve proper rendering of labels over arrows.
    (x2p) edge [bend right=25] node[left=1mm] {} (c12p)
         edge [bend right=25] node[left=1mm] {} (c22p)
         edge [bend right=0] node[left=1mm] {} (x1p)
    (x2n) edge [bend right=25] node[left=1mm] {} (c12n)
         edge [bend right=25] node[left=1mm] {} (c22n)
         edge [bend right=0] node[left=1mm] {} (x1n);
  \path[every node/.style={font=\sffamily\small,
  		fill=white,inner sep=1pt}]
  	% Right-hand-side arrows rendered from top to bottom to
  	% achieve proper rendering of labels over arrows.
    (x1p) edge [bend right=25] node[left=1mm] {} (c11p)
         edge [bend right=25] node[left=1mm] {} (c21p)
    (x1n) edge [bend right=25] node[left=1mm] {} (c11n)
         edge [bend right=25] node[left=1mm] {} (c21n);
  \path[every node/.style={font=\sffamily\small,
  		fill=white,inner sep=1pt},draw=red]
  	% Right-hand-side arrows rendered from top to bottom to
  	% achieve proper rendering of labels over arrows.
    (c1p) edge [bend right=0] node[left=1mm] {} (c2p)
          edge [bend right=0] node[left=1mm] {} (c2n)
          edge [bend right=0] node[left=1mm] {} (c2nm1p)
          edge [bend right=0] node[left=1mm] {} (c2dotsp)
          edge [bend right=0] node[left=1mm] {} (c22p)
          edge [bend right=0] node[left=1mm] {} (c21p)
          edge [bend right=0] node[left=1mm] {} (c2nm1n)
          edge [bend right=0] node[left=1mm] {} (c2dotsn)
          edge [bend right=0] node[left=1mm] {} (c22n)
          edge [bend right=0] node[left=1mm] {} (c21n)
    (c1n) edge [bend right=0] node[left=1mm] {} (c2p)
          edge [bend right=0] node[left=1mm] {} (c2n)
          edge [bend right=0] node[left=1mm] {} (c2nm1p)
          edge [bend right=0] node[left=1mm] {} (c2dotsp)
          edge [bend right=0] node[left=1mm] {} (c22p)
          edge [bend right=0] node[left=1mm] {} (c21p)
          edge [bend right=0] node[left=1mm] {} (c2nm1n)
          edge [bend right=0] node[left=1mm] {} (c2dotsn)
          edge [bend right=0] node[left=1mm] {} (c22n)
          edge [bend right=0] node[left=1mm] {} (c21n);
  \path[every node/.style={font=\sffamily\small,
  		fill=white,inner sep=1pt},draw=red]
  	% Right-hand-side arrows rendered from top to bottom to
  	% achieve proper rendering of labels over arrows.
    (c1nm1p) edge [bend right=0] node[left=1mm] {} (c2p)
          edge [bend right=0] node[left=1mm] {} (c2n)
          edge [bend right=0] node[left=1mm] {} (c2nm1p)
          edge [bend right=0] node[left=1mm] {} (c2dotsp)
          edge [bend right=0] node[left=1mm] {} (c22p)
          edge [bend right=0] node[left=1mm] {} (c21p)
          edge [bend right=0] node[left=1mm] {} (c2nm1n)
          edge [bend right=0] node[left=1mm] {} (c2dotsn)
          edge [bend right=0] node[left=1mm] {} (c22n)
          edge [bend right=0] node[left=1mm] {} (c21n)
    (c1nm1n) edge [bend right=0] node[left=1mm] {} (c2p)
          edge [bend right=0] node[left=1mm] {} (c2n)
          edge [bend right=0] node[left=1mm] {} (c2nm1p)
          edge [bend right=0] node[left=1mm] {} (c2dotsp)
          edge [bend right=0] node[left=1mm] {} (c22p)
          edge [bend right=0] node[left=1mm] {} (c21p)
          edge [bend right=0] node[left=1mm] {} (c2nm1n)
          edge [bend right=0] node[left=1mm] {} (c2dotsn)
          edge [bend right=0] node[left=1mm] {} (c22n)
          edge [bend right=0] node[left=1mm] {} (c21n);
\path[every node/.style={font=\sffamily\small,
  		fill=white,inner sep=1pt},draw=red]
  	% Right-hand-side arrows rendered from top to bottom to
  	% achieve proper rendering of labels over arrows.
    (c12p) edge [bend right=0] node[left=1mm] {} (c2p)
          edge [bend right=0] node[left=1mm] {} (c2n)
          edge [bend right=0] node[left=1mm] {} (c2nm1p)
          edge [bend right=0] node[left=1mm] {} (c2dotsp)
          edge [bend right=0] node[left=1mm] {} (c22p)
          edge [bend right=0] node[left=1mm] {} (c21p)
          edge [bend right=0] node[left=1mm] {} (c2nm1n)
          edge [bend right=0] node[left=1mm] {} (c2dotsn)
          edge [bend right=0] node[left=1mm] {} (c22n)
          edge [bend right=0] node[left=1mm] {} (c21n)
    (c12n) edge [bend right=0] node[left=1mm] {} (c2p)
          edge [bend right=0] node[left=1mm] {} (c2n)
          edge [bend right=0] node[left=1mm] {} (c2nm1p)
          edge [bend right=0] node[left=1mm] {} (c2dotsp)
          edge [bend right=0] node[left=1mm] {} (c22p)
          edge [bend right=0] node[left=1mm] {} (c21p)
          edge [bend right=0] node[left=1mm] {} (c2nm1n)
          edge [bend right=0] node[left=1mm] {} (c2dotsn)
          edge [bend right=0] node[left=1mm] {} (c22n)
          edge [bend right=0] node[left=1mm] {} (c21n);
\path[every node/.style={font=\sffamily\small,
  		fill=white,inner sep=1pt},draw=red]
  	% Right-hand-side arrows rendered from top to bottom to
  	% achieve proper rendering of labels over arrows.
    (c11p) edge [bend right=0] node[left=1mm] {} (c2p)
          edge [bend right=0] node[left=1mm] {} (c2n)
          edge [bend right=0] node[left=1mm] {} (c2nm1p)
          edge [bend right=0] node[left=1mm] {} (c2dotsp)
          edge [bend right=0] node[left=1mm] {} (c22p)
          edge [bend right=0] node[left=1mm] {} (c21p)
          edge [bend right=0] node[left=1mm] {} (c2nm1n)
          edge [bend right=0] node[left=1mm] {} (c2dotsn)
          edge [bend right=0] node[left=1mm] {} (c22n)
          edge [bend right=0] node[left=1mm] {} (c21n)
    (c11n) edge [bend right=0] node[left=1mm] {} (c2p)
          edge [bend right=0] node[left=1mm] {} (c2n)
          edge [bend right=0] node[left=1mm] {} (c2nm1p)
          edge [bend right=0] node[left=1mm] {} (c2dotsp)
          edge [bend right=0] node[left=1mm] {} (c22p)
          edge [bend right=0] node[left=1mm] {} (c21p)
          edge [bend right=0] node[left=1mm] {} (c2nm1n)
          edge [bend right=0] node[left=1mm] {} (c2dotsn)
          edge [bend right=0] node[left=1mm] {} (c22n)
          edge [bend right=0] node[left=1mm] {} (c21n);
\path[every node/.style={font=\sffamily\small,
  		fill=white,inner sep=1pt},draw=red]
  	% Right-hand-side arrows rendered from top to bottom to
  	% achieve proper rendering of labels over arrows.
    (c1dotsp) edge [bend right=0] node[left=1mm] {} (c2p)
          edge [bend right=0] node[left=1mm] {} (c2n)
          edge [bend right=0] node[left=1mm] {} (c2nm1p)
          edge [bend right=0] node[left=1mm] {} (c2dotsp)
          edge [bend right=0] node[left=1mm] {} (c22p)
          edge [bend right=0] node[left=1mm] {} (c21p)
          edge [bend right=0] node[left=1mm] {} (c2nm1n)
          edge [bend right=0] node[left=1mm] {} (c2dotsn)
          edge [bend right=0] node[left=1mm] {} (c22n)
          edge [bend right=0] node[left=1mm] {} (c21n)
    (c1dotsn) edge [bend right=0] node[left=1mm] {} (c2p)
          edge [bend right=0] node[left=1mm] {} (c2n)
          edge [bend right=0] node[left=1mm] {} (c2nm1p)
          edge [bend right=0] node[left=1mm] {} (c22p)
          edge [bend right=0] node[left=1mm] {} (c21p)
          edge [bend right=0] node[left=1mm] {} (c2nm1n)
          edge [bend right=0] node[left=1mm] {} (c2dotsn)
          edge [bend right=0] node[left=1mm] {} (c22n)
          edge [bend right=0] node[left=1mm] {} (c21n)
          edge [bend right=0] node[left=0mm] {} (c2dotsp);

  \node[main node] (xp) {$x_{n+1}^+$};
  \node[main node] (xn) [below of=xp] {$x_{n+1}^-$};
  \node[main node] (c1p) [below of=xn] {$c_{n+1}^{1+}$};
  \node[main node] (c1n) [below of=c1p] {$c_{n+1}^{1-}$};
  \node[main node] (c2p) [below of=c1n] {$c_{n+1}^{2+}$};
  \node[main node] (c2n) [below of=c2p] {$c_{n+1}^{2-}$};
  \node[main node] (xnm1p) [right of=xp] {$x_{n}^+$};
  \node[main node] (xnm1n) [right of=xn] {$x_{n}^-$};
  \node[main node] (c1nm1p) [right of=c1p] {$c_{n}^{1+}$};
  \node[main node] (c1nm1n) [right of=c1n] {$c_{n}^{1-}$};
  \node[main node] (c2nm1p) [right of=c2p] {$c_{n}^{2+}$};
  \node[main node] (c2nm1n) [right of=c2n] {$c_{n}^{2-}$};
  \node[main node] (xdotsp) [right of=xnm1p] {$\dots$};
  \node[main node] (xdotsn) [right of=xnm1n] {$\dots$};
  \node[main node] (c1dotsp) [right of=c1nm1p] {$\dots$};
  \node[main node] (c1dotsn) [right of=c1nm1n, label=below:\Huge Hebb] {$\dots$};
  \node[main node] (c2dotsp) [right of=c2nm1p] {$\dots$};
  \node[main node] (c2dotsn) [right of=c2nm1n] {$\dots$};
  \node[main node] (x2p) [right of=xdotsp] {$x_{2}^+$};
  \node[main node] (x2n) [right of=xdotsn] {$x_{2}^-$};
  \node[main node] (c12p) [right of=c1dotsp] {$c_{2}^{1+}$};
  \node[main node] (c12n) [right of=c1dotsn] {$c_{2}^{1-}$};
  \node[main node] (c22p) [right of=c2dotsp] {$c_{2}^{2+}$};
  \node[main node] (c22n) [right of=c2dotsn] {$c_{2}^{2-}$};
  \node[main node] (x1p) [right of=x2p] {$x_{1}^+$};
  \node[main node] (x1n) [right of=x2n] {$x_{1}^-$};
  \node[main node] (c11p) [right of=c12p] {$c_{1}^{1+}$};
  \node[main node] (c11n) [right of=c12n] {$c_{1}^{1-}$};
  \node[main node] (c21p) [right of=c22p] {$c_{1}^{2+}$};
  \node[main node] (c21n) [right of=c22n] {$c_{1}^{2-}$};

\end{tikzpicture}
}
\caption{\small The neural circuit that computes covariances, $\sigma_i^{++}$, $\sigma_i^{+-}$, etc., between lagged values, $x(t - T)$, of the input time series. The mean, $m_0$, is first removed from the series. Then, using PMPY pairs, positive and negative values of the series are propagated for $n + 1$ lags. These same values are copied into two sets of populations among which are feedforward, all-to-all connections. Hebbian plasticity (Hebb) acts on the synapses between the two sets of populations, generating a synaptic connectivity that is an estimate of the lagged covariance.}
\end{figure}

\subsection{Recursive, Gradient Descent Solution to $\boldsymbol{\sigma} = \Sigma \mathbf{a}$}

To make predictions, we need to estimate $\mathbf{a}$ for the process. To solve $\boldsymbol{\sigma} = \Sigma \mathbf{a}$, we find the (unique) zero of $\boldsymbol{\sigma} - \Sigma \mathbf{a}$ using gradient descent:
\begin{eqnarray}
  \tau \frac{d\mathbf{a}}{dt} & = & -\frac{1}{2}\frac{\delta}{\delta \mathbf{a}} \left( \| \boldsymbol{\sigma} - \Sigma \mathbf{a} \|^2 \right) \nonumber \\
  & = & \Sigma \left( \boldsymbol{\sigma} - \Sigma \mathbf{a} \right) \; . \nonumber
\end{eqnarray}
Discretizing to first order in time gives
\begin{equation}
  \mathbf{a}_{n+1} = \mathbf{a}_n + \frac{\Delta t}{\tau} \Sigma \left( \boldsymbol{\sigma} - \Sigma \mathbf{a}_n \right) \; , \nonumber
\end{equation}
and, since the eigenvalues of the symmetric matrix $\Sigma$ are positive and as long as $\Delta t/\tau$ is sufficiently small, iteration will give $\mathbf{a}_\infty \rightarrow \Sigma^{-1} \boldsymbol\sigma$ as $\tau \rightarrow \infty$.

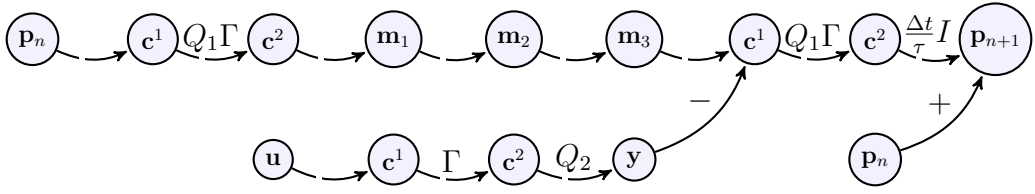
\begin{figure*}
\centering
\begin{tikzpicture}[->,>=stealth',shorten >=1pt,auto,node distance=2.0cm,
  thick,main node/.style={circle,fill=blue!5,draw,
  font=\sffamily\large\bfseries,minimum size=5mm},scale=0.80,every node/.style={transform shape}]

  \node[main node] (pna) {$\mathbf{p}_n$};
  \node[main node] (c1a) [right of=pna] {$\mathbf{c}^1$};
  \node[main node] (c2a) [right of=c1a] {$\mathbf{c}^2$};
  \node[main node] (m1)  [right of=c2a] {$\mathbf{m}_1$};
  \node[main node] (m2) [right of=m1] {$\mathbf{m}_2$};
  \node[main node] (m3) [right of=m2] {$\mathbf{m}_3$};
  \node[main node] (z) [right of=m3] {$\mathbf{c}^1$};
  \node[main node] (zp) [right of=z] {$\mathbf{c}^2$};
  \node[main node] (pnp1) [right of=zp] {$\mathbf{p}_{n+1}$};
  \node[main node] (u) [below of=c2a] {$\mathbf{u}$};
  \node[main node] (c1b) [below of=m1] {$\mathbf{c}^1$};
  \node[main node] (c2b) [below of=m2] {$\mathbf{c}^2$};
  \node[main node] (y) [below of=m3] {$\mathbf{y}$};
  \node[main node] (pnb) [below of=zp] {$\mathbf{p}_n$};

\path[every node/.style={font=\sffamily\small,
  		fill=white,inner sep=1pt, thick}]
  	% Right-hand-side arrows rendered from top to bottom to
  	% achieve proper rendering of labels over arrows.
    (pna) edge [bend right=25] node[left=1mm] {} (c1a)
    (c1a) edge [bend right=25] node[left=1mm,label=above:\large $\;\; Q_1 \Gamma$] {} (c2a)
    (z) edge [bend right=25] node[left=1mm,label=above:\large $\;\; Q_1 \Gamma$] {} (zp)
    (c2a) edge [bend right=25] node[left=1mm] {} (m1)
    (m1) edge [bend right=25] node[left=1mm] {} (m2)
    (m2) edge [bend right=25] node[left=1mm] {} (m3)
    (m3) edge [bend right=25] node[left=1mm] {} (z)
    (u) edge [bend right=25] node[left=1mm] {} (c1b)
    (c1b) edge [bend right=25] node[left=1mm,label=above:\large $\;\; \Gamma$] {} (c2b)
    (c2b) edge [bend right=25] node[left=1mm,label=above:\large $\;\; Q_2$] {} (y)
    (y) edge [bend right=25] node[left=1mm,label=above:\large $-$] {} (z)
    (pnb) edge [bend right=25] node[left=1mm,label=above:\large $+$] {} (pnp1)
    (zp) edge [bend right=25] node[left=1mm,label=above:\large $\;\; \frac{\Delta t}{\tau} I$] {} (pnp1);

\end{tikzpicture}
\caption{\small The gradient descent algorithm: the coefficient estimate, $\mathbf{p}_n$, a vector of PMPY pairs, here represented as one node in the graph, is copied via pulse gating into $\mathbf{c}^1$, the vector of PMPY pairs in the middle two rows of Fig. 1, here represented as one node. This vector is then transformed via the synaptic connectivities, $\Gamma$, learned with Hebbian plasticity, using appropriate pulsing, $Q_1$ acts on the output, with the result that $Q_1 \Gamma \mathbf{p}_n$ is written into the PMPY vector $\mathbf{c}^2$, in the lower two rows of Fig. 1. This result is then propagated through a set of memory populations, $\mathbf{m}_{1,\dots,3}$. After a sufficient delay such that $\mathbf{c}^1$ has had time to decay to a firing rate of approximately $0$, $\mathbf{u}$ is read into $\mathbf{c}^1$, then transformed such that $\Gamma \mathbf{u}$ is now in $\mathbf{c}^2$. The fixed connectivity $Q_2$ acts on this result giving $Q_2 \Gamma \mathbf{u}$ in $\mathbf{y}$. The streams are then merged with inhibitory connectivity acting to subtract $\mathbf{m}_3$ from $\mathbf{y}$, such that the value $Q_2 \Gamma \mathbf{u} - Q_1 \Gamma \mathbf{p}_n$ is in $\mathbf{z}$. Finally, via the last two operations, $\mathbf{p}_{n+1}$ is set to $\mathbf{p}_n + \frac{\Delta t}{\tau} Q_1 \Gamma \left( \boldsymbol\gamma - Q_1 \Gamma \mathbf{p}_n \right)$. In this way, one step of gradient descent is achieved.}
\end{figure*}

In order to see how this solution may be computed using positive-only elements such as when performing the calculation with a set of PMPY pairs, we consider a pair of positive, semi-definite time series derived from $x_t$, $(x^+_t, x^-_t)$, where $x^+_t - x^-_t = x_t$. That is, the part of $x_t$ that is greater than the mean in a PMPY pair is contained in $x^+_t$, and the absolute value of the part of $x_t$ that is less than the mean is contained in $x^-_t$. The covariances associated with this time series are $\sigma^{++}_i \equiv \langle x^+_{t-i} x^+_t \rangle$, $\sigma^{+-}_i \equiv \langle x^+_{t-i} x^-_t \rangle$, $\sigma^{-+}_i \equiv \langle x^-_{t-i} x^+_t \rangle$, and $\sigma^{--}_i \equiv \langle x^-_{t-i} x^-_t \rangle$. Note that $\sigma_i = \sigma^{++}_i - \sigma^{+-}_i - \sigma^{-+}_i + \sigma^{--}_i$. Note also that if we define
$$\Xi = \left[ \begin{array} {cccccccc} 
               1 & -1 & 0 & 0 & 0 & 0 & \dots & \dots \\ 
               0 & 0 & 1 & -1 & 0 & 0 & \dots & \dots \\
               0 & 0 & 0 & 0 & 1 & -1 & \dots & \dots \end{array} \right] \; ,$$
\begin{equation}\Gamma = \left[ \begin{array} {cccccccc}
               \sigma_0^{++} & \sigma_0^{+-} & \sigma_1^{++} & \sigma_1^{+-} & \dots & \dots \\
               \sigma_0^{-+} & \sigma_0^{--} & \sigma_1^{-+} & \sigma_1^{--} & \dots & \dots \\
               \sigma_1^{++} & \sigma_1^{+-} & \sigma_0^{++} & \sigma_0^{+-} & \dots & \dots \\
               \sigma_1^{-+} & \sigma_1^{--} & \sigma_0^{-+} & \sigma_0^{--} & \dots & \dots \\
               \vdots & \vdots & \vdots & \vdots & \ddots & \ddots \\
               \vdots & \vdots & \vdots & \vdots & \ddots & \ddots \end{array} \right] \label{biggamma}
\end{equation}
$\boldsymbol{\gamma} = (\sigma_1^{++}, \sigma_1^{+-}, \sigma_2^{++}, \sigma_2^{+-}, \dots)$, and $\mathbf{p} = (a_1^+, a_1^-, a_2^+, a_2^-, \dots)$, then
$\Xi \boldsymbol{\gamma} = \boldsymbol{\sigma}$, and $\Xi \Gamma \Xi^T = \Sigma$. Thus, if we first compute the solution to
$$\tau \frac{d\mathbf{p}}{dt} = \Gamma \left( \boldsymbol{\gamma} - \Gamma \mathbf{p} \right)$$
then, $\Xi \mathbf{p} = \mathbf{a}$. It may be seen directly that this gradient descent projects to the one described above since
\begin{eqnarray}
  \tau \frac{d\mathbf{a}}{dt} = \tau \Xi \frac{d\mathbf{p}}{dt} & = & \Xi \Gamma \Xi^T \left( \Xi \boldsymbol{\gamma}   
    - \Xi \Gamma \Xi^T \Xi \mathbf{p} \right) \nonumber \\
    & = & \Sigma \left( \boldsymbol{\sigma} - \Sigma \mathbf{a} \right) \; .
\end{eqnarray}
The reason this works is that multiplication of a vector, $\mathbf{v}$, by $\Xi$ returns a new vector containing the differences of each pair of elements of $\mathbf{v}$, and the matrix product $\Xi \mathbf{X} \Xi^T$ returns a new matrix containing the difference of the sum of diagonal elements of each $2 \times 2$ block submatrix and the sum of its off diagonal elements.

Finally, it will be useful in the discussion of the neural circuit used to implement gradient descent to extend the above considerations by using a matrix $\Gamma'$, which is defined as in (\ref{biggamma}), but based on an $(n + 1) \times (n + 1)$ matrix $\Sigma$. Thus, $\Gamma'$ will contain covariances up to a lag of $n + 1$. In this case, we can explicitly write the gradient descent algorithm as
\begin{equation}
  \mathbf{p}_{n+1}' = \mathbf{p}_n' + \frac{\Delta t}{\tau} Q_1 \Gamma' \left( \boldsymbol\gamma' - Q_1 \Gamma' \mathbf{p}_n' \right) \; , \label{pprime}
  \end{equation}
where $\boldsymbol\gamma' \equiv Q_2 \Gamma' \mathbf{u}$, $\mathbf{u} \equiv (1, 0, 0, 0, \dots)$, 
\begin{eqnarray}
  Q_1 & = & \mathrm{diag}([1, 1, \dots, 1, 1, 0, 0]) \nonumber \\
  Q_2 & = & \mathrm{diag}([1, 1, \dots, 1, 1]_n, 2) \nonumber
\end{eqnarray}
(i.e. $Q_2$ is a matrix of zeros, except for a $2$nd superdiagonal of $1$'s), and $$\mathbf{p}' = (a_1^+, a_1^-, \dots, a_n^+, a_n^-, 0, 0) \; .$$

We have done this seemingly nonsensical extension because now $\gamma'$ is generated by a vector input to the covariance matrix $\Gamma'$, as opposed to being a fixed vector. Thus, the algorithm will work for an arbitrary AR(n) time series covariance.

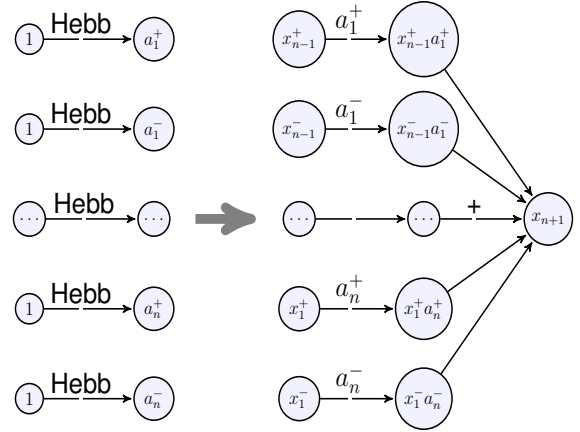
\begin{figure}
\centering
\resizebox{3in}{2.4in}{
\begin{tikzpicture}[->,>=stealth',shorten >=1pt,auto,node distance=3.0cm,
  thick,main node/.style={circle,fill=blue!5,draw,
  font=\sffamily\large\bfseries,minimum size=5mm},scale=0.8,every node/.style={transform shape}]

  \node[main node] (1a) {${1}$};
  \begin{scope}[node distance = 2.0cm]
  \node[main node] (1b) [below of=1a] {${1}$};
  \node[main node] (dotsa) [below of=1b] {$\dots$};
  \node[main node] (1c) [below of=dotsa] {${1}$};
  \node[main node] (1d) [below of=1c] {${1}$};
  \end{scope}
  \node[main node] (aa) [right of=1a] {${a}^+_1$};
  \node[main node] (ab) [right of=1b] {${a}^-_1$};
  \node[main node] (dotsb) [right of=dotsa] {$\dots$};
  \node[main node] (ac) [right of=1c] {${a}^+_n$};
  \node[main node] (ad) [right of=1d] {${a}^-_n$};

  \path[every node/.style={font=\sffamily\small,
  		fill=white,inner sep=1pt, thick}]
  	% Right-hand-side arrows rendered from top to bottom to
  	% achieve proper rendering of labels over arrows.
    (1a) edge [bend right=0] node[left=1mm, label=above:\Large Hebb] {} (aa)
    (1b) edge [bend right=0] node[left=1mm, label=above:\Large Hebb] {} (ab)
    (dotsa) edge [bend right=0] node[left=1mm, label=above:\Large Hebb] {} (dotsb)
    (1c) edge [bend right=0] node[left=1mm, label=above:\Large Hebb] {} (ac)
    (1d) edge [bend right=0] node[left=1mm, label=above:\Large Hebb] {} (ad);

  \begin{scope} [node distance = 3.5cm]
  \node[main node] (2a) [right of=aa] {${x_{n-1}^+}$};
  \end{scope}
  \begin{scope}[node distance = 2.0cm]
  \node[main node] (2b) [below of=2a] {${x_{n-1}^-}$};
  \node[main node] (2dotsa) [below of=2b] {$\dots$};
  \node[main node] (2c) [below of=2dotsa] {${x_1^+}$};
  \node[main node] (2d) [below of=2c] {${x_1^-}$};
  \end{scope}
  \node[main node] (2aa) [right of=2a] {$x_{n-1}^+ {a}^+_1$};
  \node[main node] (2ab) [right of=2b] {$x_{n-1}^- {a}^-_1$};
  \node[main node] (2dotsb) [right of=2dotsa] {$\dots$};
  \node[main node] (2ac) [right of=2c] {${x_1^+ {a}^+_n}$};
  \node[main node] (2ad) [right of=2d] {${x_1^- {a}^-_n}$};

  \path[every node/.style={font=\sffamily\small,
  		fill=white,inner sep=1pt, thick}]
  	% Right-hand-side arrows rendered from top to bottom to
  	% achieve proper rendering of labels over arrows.
    (2a) edge [bend right=0] node[left=1mm, label=above:\Large ${a}^+_1$] {} (2aa)
    (2b) edge [bend right=0] node[left=1mm, label=above:\Large ${a}^-_1$] {} (2ab)
    (2dotsa) edge [bend right=0] node[left=1mm, label=above:\Large ] {} (2dotsb)
    (2c) edge [bend right=0] node[left=1mm, label=above:\Large ${a}^+_n$] {} (2ac)
    (2d) edge [bend right=0] node[left=1mm, label=above:\Large ${a}^-_n$] {} (2ad);
    
    \node[main node] (tplus) [right of=2dotsb] {$x_{n+1}$};  

  \path[every node/.style={font=\sffamily\small,
  		fill=white,inner sep=1pt, thick}]
  	% Right-hand-side arrows rendered from top to bottom to
  	% achieve proper rendering of labels over arrows.
    (2aa) edge [bend right=0] node[left=1mm] {} (tplus)
    (2ab) edge [bend right=0] node[left=1mm] {} (tplus)
    (2dotsb) edge [bend right=0] node[left=1mm, label=above:\Large +] {} (tplus)
    (2ac) edge [bend right=0] node[left=1mm] {} (tplus)
    (2ad) edge [bend right=0] node[left=1mm] {} (tplus);

   \begin{scope} [line width=5pt]
     \useasboundingbox (0,-1.5) rectangle (3.5,1.5);
     \draw[gray,->]    (4.0,-4) -- (5.5,-4);
  \end{scope}

\end{tikzpicture}
}
\caption{\small Prediction: On the left, populations initialized with the value $1$ are connected one-to-one with populations containing estimated coefficients. Hebbian plasticity causes the values of the estimated coefficients to be encoded in the synapses. On the right, once the synapses are sufficiently stable, lagged time series values are written into the populations previously initialized with $1$ ($1$-populations). With a pulse, the positive components are computed, then summed, resulting in a prediction of $x(t+1)^+$. Initializing the $1$-populations with swapped $+$ and $-$ components (not shown) of $x$ results in $x(t+1)^-$. Estimates of future values of $x$ may subsequently be made with the same circuit.}
\end{figure}

\subsection{The Neural Circuit}

Throughout our circuits, we use the pulse-gated propagation of firing rates described in \cite{SornborgerWangTao}.

\subsubsection{Computing Lagged Covariances}

The first major structure in the circuit is set up to use Hebbian plasticity to calculate covariances, $\sigma_i^{++}$, $\sigma_i^{+-}$, etc., between successively delayed random variables in the input. To do this, the input, $x(t)$, is discretized and bound via a pulse into the circuit as current or firing rate amplitude packets, $x^+_i \equiv [x(t_i) - m_0]^+$ and $x^-_i \equiv [m_0 - x(t_i)]^+$, in PMPY pairs. These packets are propagated in a chain of $n+1$ pairs. Additionally, two copies of each pair are made using a pulse gated copying procedure.

\begin{figure*}
\centering
\includegraphics[width=6.5in]{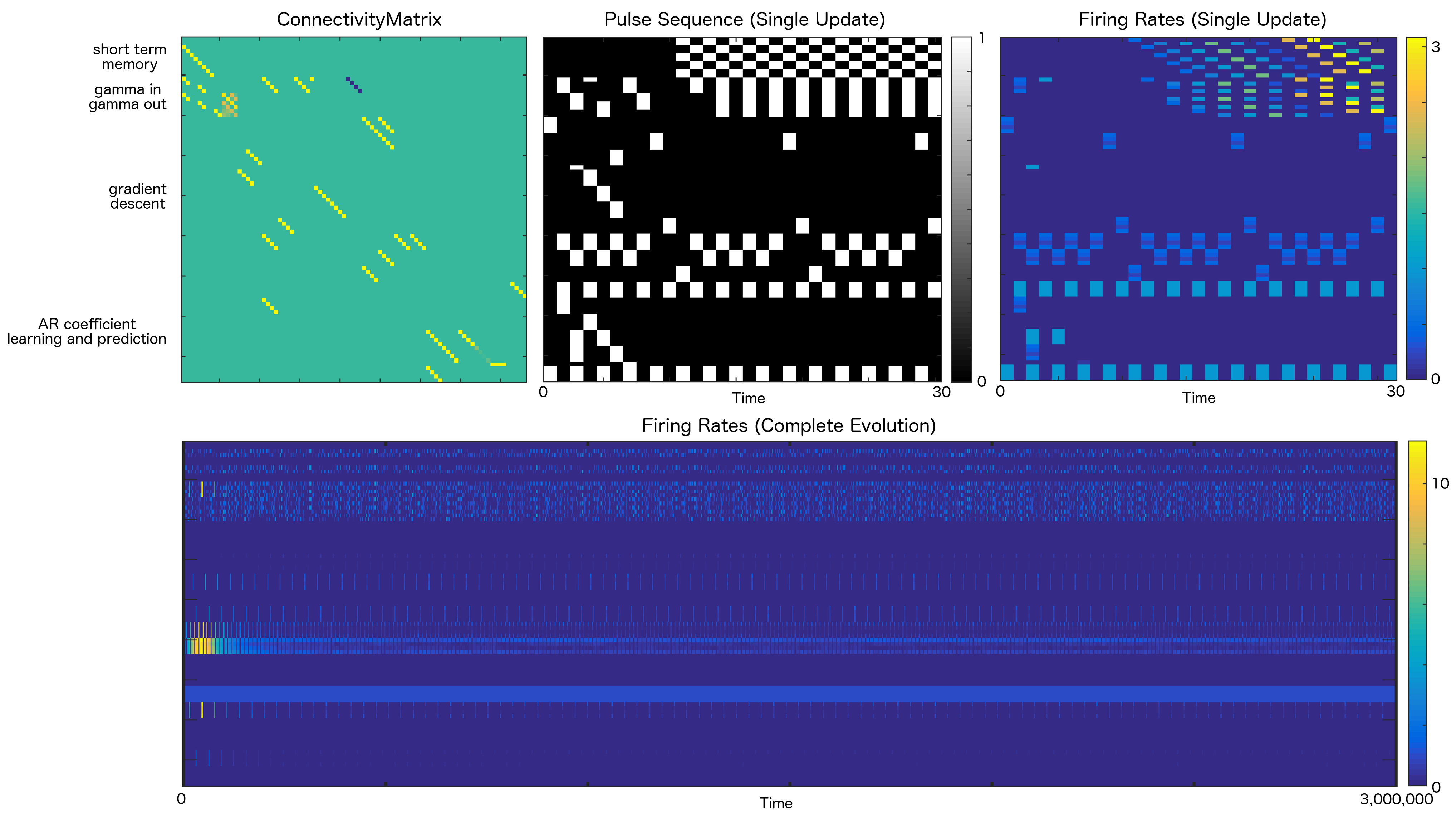}
\caption{The Basic Pulse-Gated Algorithm: Connectivity matrix (top left), pulse sequence (top center) and neuronal population firing rates (top right for a single update and bottom for the complete evolution). Covariance, gradient descent, and prediction circuits are activated such that they do not interfere with each other, see text for complete description.}
\label{PulseSequence}
\end{figure*}

Hebbian plasticity is meant to mimic synaptic plasticity. It is a phenomenon in which coincident spiking activity between two neurons serves to increase the synaptic weight between the neurons. Since the probability of two spikes being coincident is proportional to the product of the firing rates, over time the synaptic strength becomes an estimate of the covariance between the firing rates of the two neurons (or in our case, neuronal populations). In our AR circuit, the populations of copied PMPY pairs interact with all-to-all connectivity (i.e. both populations in each pair in the first set of copies are connected with both populations in each pair in the second set of copies, see Fig. 1).

Synaptic weights, $\{ s_j \}$, are set according to $$\tau_s \frac{ds_j}{dt} = - (s_j - x(t) y(t)) \; ,$$ where $x$ and $y$ are firing rates from two neuronal populations. We assume that the input is in the form of a train of delta functions, $z(t) = \sum_i z_i \delta(t - t_i)$, that are read in by a pulse-gated neural population. After gating into a neural population, we have $x(t) = z(t) \ast G(t)$, where $G(t)$ is the pulse envelope from the gating operation, $$G(t) = \left\{ \begin{array}{cc} \frac{t}{\tau} e^{-t/\tau}, & 0 < t < T\\ \frac{T}{\tau} e^{-t/\tau}, & T < t < \infty \end{array} \right. \; .$$ In the limit that the Hebbian timescale is much greater than the synaptic timescale, $\tau_s \gg \tau$, and the gated pulses overlap only negligibly, we have $$s_j(t) \approx  \hat{\sigma}_j \; ,$$ where $\hat{\sigma}_j = 4\tau/\tau_s \sum_i z_i z_{i-j}$ is an estimate of the covariance between $x$ and $y$ over $\tau_s/4 \tau$ samples.

Once the synaptic strengths in the copy populations have reached an equilibrium, and using pulse gating such that the synapses are feedforward between PMPY copy $1$ and PMPY copy $2$, we have $\mathbf{c}^2 = \Gamma \mathbf{c}^1$. Here, for simplicity, we have discarded the $'$ and just write $\Gamma$.

\subsubsection{Gradient Descent}

To implement the gradient descent part of the algorithm, we implement (\ref{pprime}) using pulse gating. A method to do this is shown in Fig. 2. Here, a vector of PMPY pairs containing $\mathbf{p}_n$ is gated into the first copy of lagged values from the time series. The synaptic connectivity, learned using Hebbian plasticity, that contains an estimate of the covariance matrix is used (along with appropriate pulses) to perform the operation $\mathbf{c}^2 := Q_1 \Gamma \mathbf{p}_n$. This result is then stored in a short-term memory. During this calculation, but after sufficient delay that $\mathbf{c}^1$ may be reused, $\mathbf{u}$ is gated into $\mathbf{c}^1$ and, via the matrix $\Gamma$ and another connectivity matrix that encodes $Q_2$, $Q_2 \Gamma \mathbf{u}$ is computed and stored in $\mathbf{y}$. The two computation streams are then combined with appropriate inhibition to subtract $\mathbf{m}_3$ from $\mathbf{y}$ giving $\mathbf{z} = \gamma - Q_1 \Gamma \mathbf{p}_n$. Two subsequent operations result in $\mathbf{p}_{n+1}$ being set to $\mathbf{p}_n + \frac{\Delta t}{\tau} Q_1 \Gamma \left( \boldsymbol\gamma - Q_1 \Gamma \mathbf{p}_n \right)$

\subsubsection{Prediction}

To predict future values of $x(t)$, we need to form estimates of $x(t)$ using the AR(n) process coefficients. The computation to do this requires us to compute $$\langle x_t^+ \rangle = \sum_{i=1}^n (\langle a_i^+ x_{t-i}^+\rangle + \langle a_i^- x_{t-i}^- \rangle)$$ and $$\langle x_t^- \rangle = \sum_{i=1}^n (\langle a_i^+ x_{t-i} ^- \rangle + \langle a_i^- x_{t-i}^+ \rangle )$$ using the coefficients, $\mathbf{p} = (a_1^+, a_1^-, \dots)$, that we have estimated with the gradient descent algorithm. One way that this may be done is depicted in Fig. 3.

Here, by initializing a set of populations to $1$ ($1$-populations), connecting them one-to-one with the elements of $\mathbf{p}_n$, Hebbian plasticity causes the synapses to encode the values of the elements of $\mathbf{p}$. This is a method for creating a long-term memory. Subsequently, $\mathbf{x}$ is copied to the $1$-populations, then gated to compute $\{ \langle a_i^+ x_{t-i}^+ \rangle, \langle a_i^- x_{t-i}^- \rangle, \dots \}$. These values are then summed to compute $\langle x_t^+ \rangle$. By swapping the $+$ and $-$ elements of $\mathbf{x}$, then copying to the $1$-populations, the same circuit results in $\langle x_t^- \rangle$. The output may be stored in memory or used in other sub-circuits in the usual way.

\section{Results}

In Fig. 4, we depict results for the $AR(2)$ process, $x_{n+1} = a_1 x_n + a_2 x_{n-1} + \epsilon$, where $\epsilon$ is a zero mean Gaussian noise process, $a_1 = 3/4$ and $a_2 = -1/2$. At the beginning of the sequence, $AR$ input to short term memory is silenced to allow operation of the gradient descent circuit. At the same time, the process coefficients, encoded in firing rate amplitudes are propagated to the coefficient learning and prediction circuit to be encoded in synaptic weights. Once the gradient descent circuit and learning and prediction circuits have updated the process coefficient estimates, $a_1$ and $a_2$, information needed for memory is propagated internally to each circuit and $AR$ input is allowed to enter the covariance learning circuit. This update sequence is then repeated to obtain progressively better estimates over a long time ($100,000$ repeats for this simulation).

\begin{figure}[!t]
\centering
\includegraphics[width=2.7in]{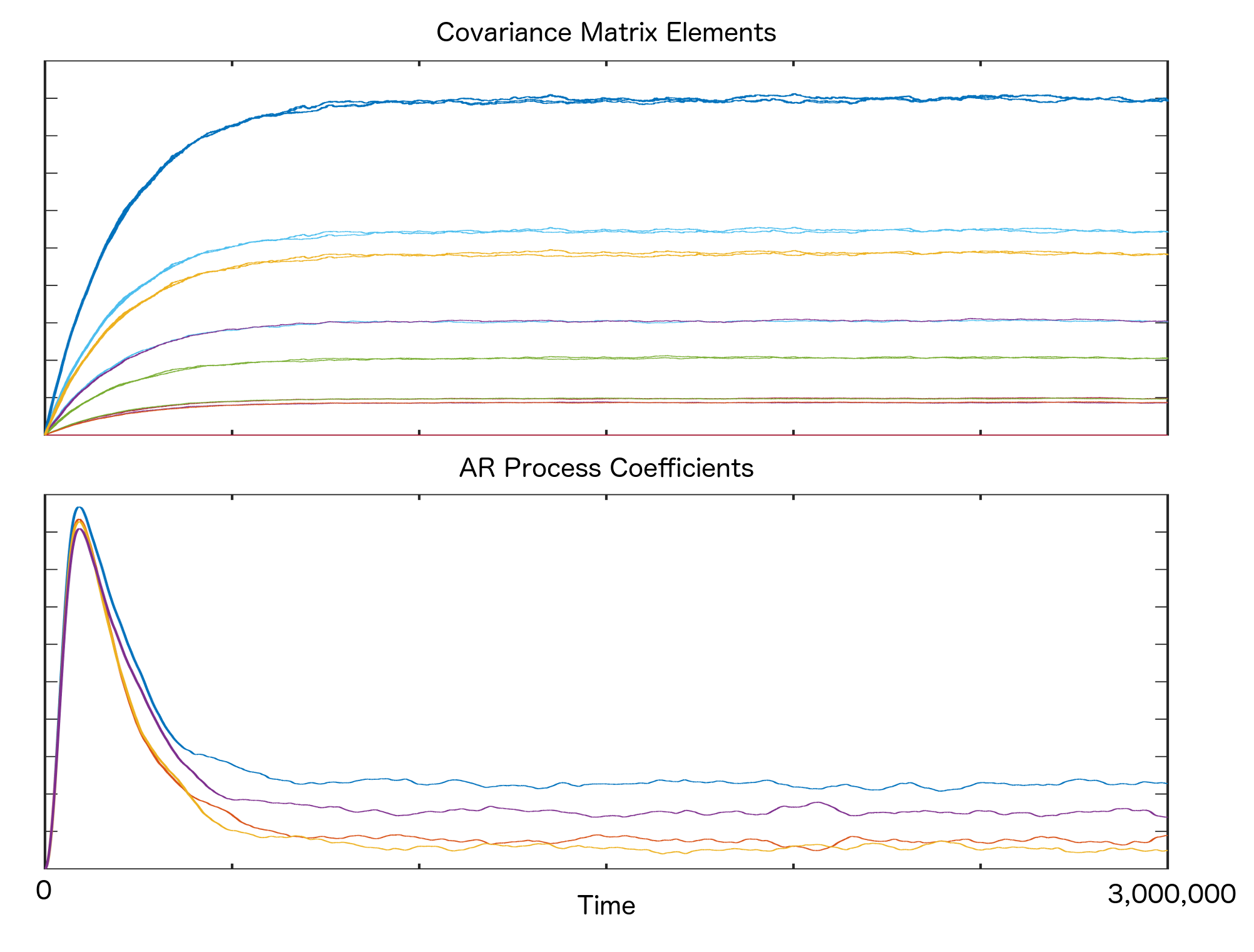}
\caption{Covariance and $AR$ Coefficient Estimates: The evolution of estimates converges after approximately $30,000$ updates (30 pulses per update).}
\label{PulseSequence}
\end{figure}

In Fig. 5, we show the convergence of the algorithm to steady state estimates for both the covariance matrix elements and $AR$ process coefficients. Final $AR$ process coefficients are within $6\%$ of theoretical values. Better estimates may be obtained by increasing the learning time constant, $\tau$.

\begin{figure}[!t]
\centering
\includegraphics[width=3.0in]{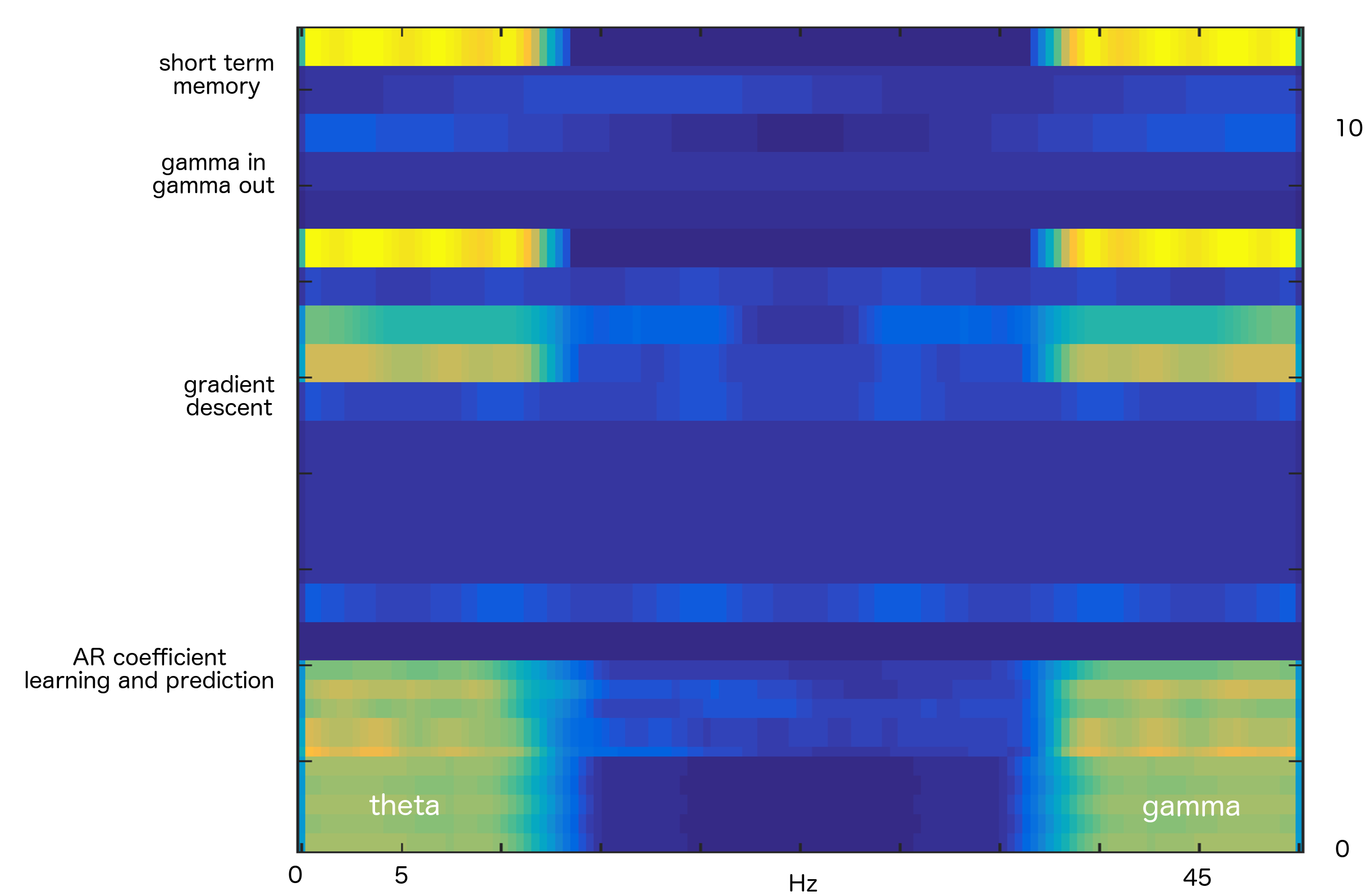}
\caption{Spectrum of $AR$ Pulse Sequence: For gating pulses of length $10$ms, the pulse sequence spectrum has peaks in the theta and gamma bands.}
\label{PulseSequence}
\end{figure}

\section{Conclusion}

Using a pulse-gating paradigm for propagating information in a neural circuit in concert with Hebbian learning, we have described how to implement an algorithm for predicting future values of an arbitrary $AR(n)$ process. We have implemented the algorithm for an $AR(2)$ process. The algorithm consists of three sub-circuits responsible for 1) learning the covariance matrix, 2) computing the process coefficients and 3) making a prediction by learning the process coefficients, then performing the prediction.

Using only pulse-gating, Hebbian learning and standard neuronal synaptic properties, we implemented a short-term memory, a gradient descent algorithm, a long-term memory and a method for computing an inner product to make a prediction. Additionally, the structure of the algorithm defines the brain rhythms that arise during information processing. In Fig. 6, we show the spectra of the gating pulses in the pulse sequence. For this figure, we used pulses of $10$ms duration, resulting in strong gamma and theta rhythms. To our knowledge, this is the first time a neuronal model has been used to relate algorithmic structure with neuronal oscillation structure.

\section*{Acknowledgment}
L.T. thanks the UC Davis Mathematics Department for its hospitality. A.T.S. would like to thank Liping Wei and the Center for
Bioinformatics at the College of Life Sciences at Peking University for their hospitality. This work was supported by the Natural Science Foundation of China grant 91232715 (YXS and LT), by the Open Research Fund of the State Key Laboratory of Cognitive Neuroscience and Learning grant CNLZD1404 (YXS and LT), and by the Beijing Municipal Science and Technology Commission under contract Z151100000915070 (YXS and LT).



%\bibliographystyle{IEEEtran}
\bibliographystyle{IEEEtran}
\bibliography{biblio}

\end{document}